\documentclass[twocolumn]{aastex631}

\usepackage{soul}
\usepackage{comment}
\usepackage{amsmath}
\usepackage{amssymb}
\usepackage{bbold}
\usepackage{hyperref}
\usepackage[super]{nth}
\usepackage[normalem]{ulem}

\newcommand\bb[1]{\mbox{\boldmath{$#1$}}}
\newcommand\bdbldot{\,\bb{:}\,}

\begin{document}

\title{Stability Analysis of the Proton Hammerhead Distribution Observed by Parker Solar Probe: Linear Theory and Fully Kinetic Simulations under Idealised Conditions}

\author[0000-0001-5079-7941]{L. Pezzini}
\affiliation{Centre for mathematical Plasma Astrophysics, Department of Mathematics, KU Leuven, Celestijnenlaan 200B, B-3001 Leuven, Belgium}
\affiliation{Solar-Terrestrial Centre of Excellence--SIDC, Royal Observatory of Belgium, Ringlaan 3 Avenue Circulaire, B-1180 Brussels, Belgium}
\author[0000-0003-1138-652X]{J. L. Verniero}
\affiliation{Code 672, NASA, Goddard Space Flight Center, Greenbelt, MD 20771, USA}
\author[0000-0002-7526-8154]{F. Bacchini}
\affiliation{Centre for mathematical Plasma Astrophysics, Department of Mathematics, KU Leuven, Celestijnenlaan 200B, B-3001 Leuven, Belgium}
\affiliation{Royal Belgian Institute for Space Aeronomy, Solar-Terrestrial Center of Excellence--SIDC, Ringlaan 3 Avenue Circulaire, B-1180 Brussels,
Belgium}
\author[0000-0002-2542-9810]{A. N. Zhukov}
\affiliation{Solar-Terrestrial Centre of Excellence--SIDC, Royal Observatory of Belgium, Ringlaan 3 Avenue Circulaire, B-1180 Brussels, Belgium}
\affiliation{Skobeltsyn Institute of Nuclear Physics, Moscow State University, 119991 Moscow, Russia}
\author[0000-0001-7233-2555]{G. Arrò}
\affiliation{Department of Physics, University of Wisconsin-Madison, Madison, WI 53706, USA}
\author[0000-0003-3223-1498]{R. A. López}
\affiliation{Research Center in the intersection of Plasma Physics, Matter, and Complexity ($P^2 mc$),\\ Comisi\'on Chilena de Energ\'{\i}a Nuclear, Casilla 188-D, Santiago, Chile}
\affiliation{Departamento de Ciencias F\'{\i}sicas, Facultad de Ciencias Exactas, Universidad Andres Bello, Sazi\'e 2212, Santiago 8370136, Chile}
\begin{abstract}

The non-adiabatic heating of the slow Solar Wind (SW) remains an open problem, with wave--particle interactions as a primary candidate mechanism. Novel in situ Parker Solar Probe (PSP) observations reveal strongly perpendicular anisotropic velocity distribution functions (VDFs), called ``hammerhead'', correlated with intense wave activity. These VDFs are systematically measured at the Heliospheric Current Sheet (HCS), making the hammerhead an important kinetic signature of the slow SW. In this work, we employ a fully kinetic particle-in-cell approach, complemented by a linear Vlasov solver to cross-validate the simulation results, to investigate the stability of these VDFs, the timescales over which they evolve, and their interaction with plasma waves. Our main findings indicate that the hammerhead distribution is primarily susceptible to drift-type instabilities, while energy is nonlinearly transferred back to the plasma through a combination of Landau and cyclotron resonances, resulting in net heating in the parallel direction, preferentially energizing the beam proton population. Crucially, these nonlinear interactions do not drastically alter the morphology of the distribution. This suggests the possibility that the hammerhead may be generated locally within the HCS in the inner heliosphere and subsequently advected outward, where it is eventually measured by PSP. This work lays the ground for future investigations into the kinetic physics of the HCS.

\end{abstract}


\section{Introduction} \label{sec:intro}

%
%
The thermally unstable solar corona continuously emanates a supersonic outflow of magnetized plasma into interplanetary space \citep{parker1958}, first detected by \citet{gringauz1960} using data from the Soviet Luna spacecraft, confirmed shortly after by in situ measurements from the Mariner~2 mission \citep{neugebauer1962}, and known today as the SW (see e.g., review by \citep{verscharen2019a}). The bipolar component of the solar magnetic field, stretched by the solar wind, gives rise to the HCS \citep{smith2001}, a site where magnetic reconnection drives particle energization \citep{gosling2012, phan2021}. 

%
%
The SW is fundamentally a cross-scale system where energy is cascading from large scales to small scales, where it is rearranged \citep{bruno2013}. Macroscopically, the SW expands radially from the Sun, yet its observed temperature profiles \citep{gazis1982} decay more slowly than predicted by double-adiabatic theory \citep{chew1956}. Microscopically, its low collisionality yields departures from local thermodynamic equilibrium (LTE) \citep{spitzer1962, marsch1982b, marsch2006}.  

%
%
At kinetic scales, departures from LTE, manifesting a breaking of Maxwellian symmetry, drive plasma instabilities \citep{gary1993} that release energy through wave amplification \citep{stix1962}, which back-reacts on the plasma via resonant wave--particle interactions, including Landau damping \citep{landau1946, tenbarge2013}, ion-cyclotron resonance \citep{hollweg2002a}, and transit-time damping \citep{barnes1966}. These breakings of Maxwellian symmetry appear as temperature anisotropies and particle beams (Strahl for electrons), observed in both electrons \citep{pilipp1987, bonhome2026, micera2020, micera2020a, micera2021, verscharen2022, verscharen2022a, verscharen2026} and ions \citep{tu2004, marsch2006}. Ion beams in the solar wind have been extensively studied since the early in situ observations at heliocentric distances $R \gtrsim 0.3$~au \citep{marsch1982b, podesta2011a, verscharen2013, alterman2018}, yet their origin at $R < 0.3$~au remains poorly constrained, motivating the Parker Solar Probe (PSP) mission \citep{fox2016}, whose Solar Wind Electrons Alphas and Protons (SWEAP; \citealt{kasper2016}) and FIELDS \citep{bale2016} instrument suites provide unprecedented in situ measurements in the near-Sun environment.

%
%
Ion-scale waves are a ubiquitous feature of the solar wind, with power spectra cutting off at the proton gyration scale, as also seen in the PSP encounter data considered here \citep{bowen2020, verniero2020, klein2021, verniero2022}. The strong wave activity, dominated by a few nearly monochromatic waves, as reflected in the narrowband, peaked wave spectrum, may indicate that PSP was flying through a quasi-laminar stream of plasma \citep{klein2021, verniero2022}. In these conditions, proton distributions, and in general ions, are often measured out of Maxwellian equilibrium, with VDFs well approximated by two bi-Maxwellian components with a relative drift speed: a denser core and a faster beam \citep{klein2021}. Novel PSP observations revealed proton phase-space distributions approximated by a three-component bi-Maxwellian, where a third population was needed to capture a faster population undergoing perpendicular velocity-space diffusion, dubbed the hammerhead distribution \citep{verniero2022}. A number of studies have been conducted to understand the formation mechanism and evolution of this previously unobserved structure. Analytical calculations propose that the hammerhead distribution originates at interchange reconnection sites in the low corona, where plasmas from coronal holes and closed magnetic loops mix together \citep{krasnoselskikh2023}. Linear theory shows hammerhead-like features arising from phase-space diffusion driven by a super-Alfv\'{e}nic drifting proton beam \citep{shaaban2024}. Data-constrained kinetic simulations show the impact of wave--particle resonant interactions on proton heating, reshaping the VDF \citep{klein2021, pezzini2024, lopez2026}, as well as the role of $\text{He}^{2+}$ particles \citep{ofman2025, pezzini2026}. 

A growing body of work suggests that ion-cyclotron waves are more efficient than broadband turbulence at producing hammerhead-like VDF features \citep{gonzalez2024, malaspina2024, larosa2025}. Observations of reconnection exhausts in the HCS \citep{lavraud2020, fargette2026} during PSP encounters 08 and 07 revealed energised protons leaking out of the exhaust along separatrix field lines, forming field-aligned energetic beams accompanied by strahl electrons \citep{phan2022}. In addition, a more systematic analysis employing the \textsc{hampy} convolution-based hammerhead detection method \citep{bharatidas2026a} demonstrates that hammerhead occurrences peak around HCS crossings. In current sheets, rotational discontinuities are present, such as those in switchback boundaries, and play a crucial role in generating the proton beam in SW plasma in the inner heliosphere \citep{lin2026}.
%
%
This paper presents a stability analysis of the hammerhead VDF. Using the measured parameters from \citet{verniero2022}, linear theory and nonlinear PIC simulations are combined to address the evolution of the hammerhead VDF; an estimate of the lifespan of the distribution is further provided in order to constrain its possible origin site. The remainder of this paper is organised as follows. Section~\ref{sec:linear} discusses the linear theory of the unstable eigenmodes developing from the initial conditions, with particular emphasis on the reduced mass ratio. Section~\ref{sec:setup} describes the numerical setup of the PIC simulation, including the initial conditions, the modeling hypotheses, and the kinetic and numerical parameters, the latter chosen to ensure stability and accurate resolution of the relevant physical scales. Section~\ref{sec:results} is divided into the following subsections. Subsection~\ref{subsec:phases} illustrates the global phases through which the system evolves. Subsection~\ref{subsec:wave} studies the spectral properties of the system and compares them with the results from linear theory. Subsection~\ref{subsec:vdf} presents an extensive analysis of phase space, supplemented by temporal series of the distribution function. Subsection~\ref{subsec:fpc} characterises the damping mechanisms through the field-particle correlation analysis. Finally, Section~\ref{sec:discludion} provides a comprehensive discussion and conclusions.

\section{Linear analysis}\label{sec:linear}

\begin{figure*}[ht!]
\includegraphics[width=1\textwidth]{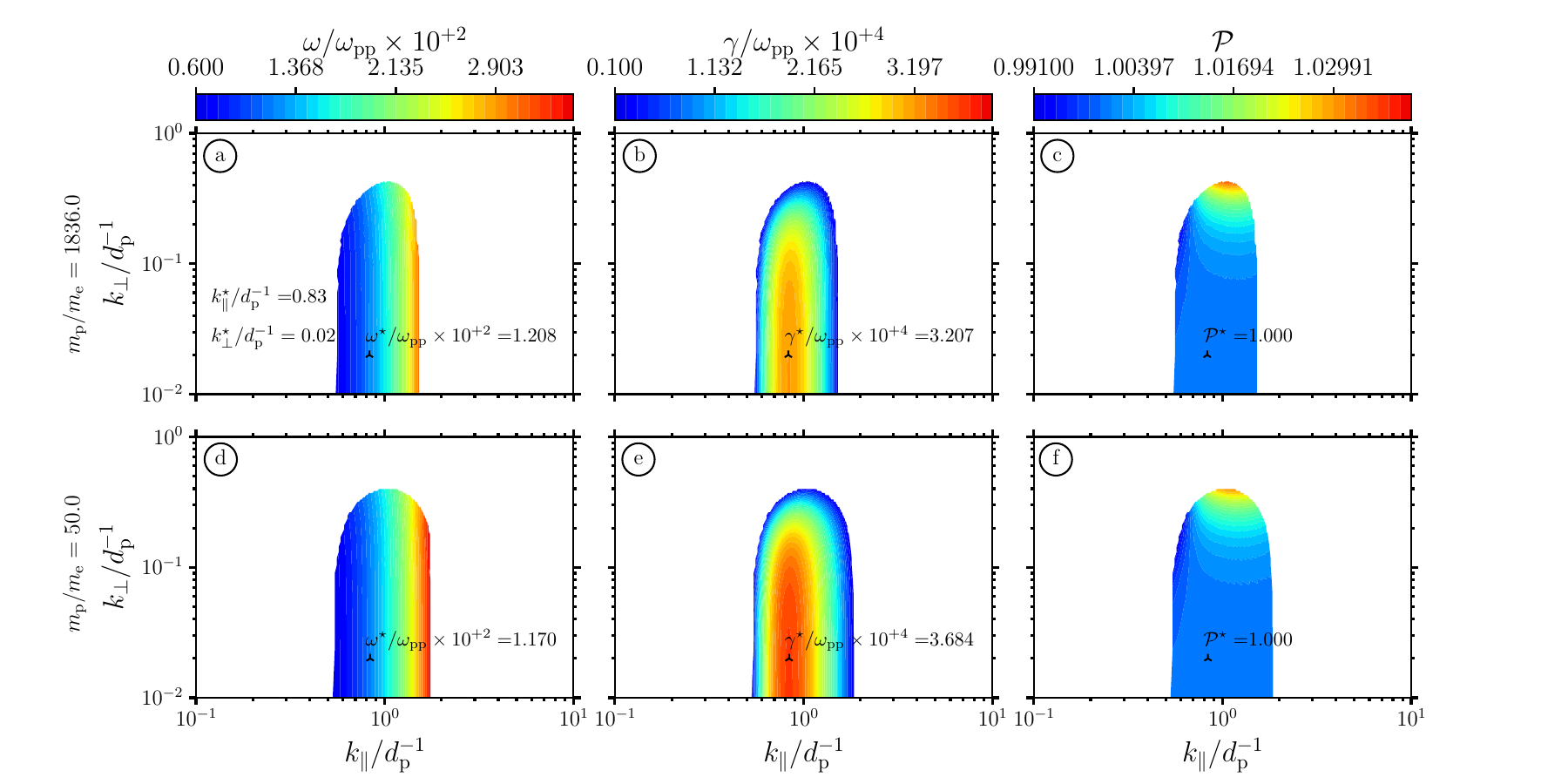}
\caption{Eigenmode spectra of the unstable system in $k_{\parallel}$--$k_{\perp}$ space. Each column shows, from left to right, the real frequency $\omega/\omega_{\mathrm{pp}}$, the growth rate $\gamma/\omega_{\mathrm{pp}}$, and the wave polarization $\mathcal{P}$, computed for a realistic mass ratio $m_p/m_e = 1836$ (top row) and a reduced one $m_p/m_e = 50$ (bottom row). The black marker indicates the position of the most unstable eigenmode.}
\label{fig:qlt}
\end{figure*}

Linear analysis of the Vlasov--Maxwell system provides a first-order approximation to the dispersion relation of the non-equilibrium plasma under consideration \citep{verniero2022}, yielding the real and imaginary parts of the frequency as a function of the wavevector $\boldsymbol{k}$. In this work, we employ the \texttt{DIS-K}\footnote{Publicly available at \url{https://github.com/ralopezh/dis-k}.} linear solver \citep{lopez2021, lopez2023} for two purposes: first, to determine the optimal parameter setup for the fully kinetic nonlinear simulation (see Section~\ref{sec:setup}); second, to benchmark the simulation against linear theory during its quasi-linear phase (see Section~\ref{sec:results}), thereby validating the quantitative reliability of our results.

The background magnetic field at the initialisation is uniform and oriented along the $x$-direction, $\boldsymbol{B}_{0} = B_0\,\hat{\boldsymbol{e}}_{x}$. The $x$-axis defines the parallel direction with respect to $\boldsymbol{B}_0$, while the $y$- and $z$-axes represent the two perpendicular directions. Hereafter, subscripts $\parallel$ and $\perp$ implicitly refer to the field-aligned frame.
Electrons are initialized in a Maxwellian equilibrium, while protons are initialized with a bi-Maxwellian distribution drifting along the parallel direction. The distribution function for species $s$ is defined as
\begin{equation}
    \begin{aligned}
        f_{\mathrm{bi\text{-}M},\, s}\!\left(v_{\parallel}, v_{\perp}\right) 
        \doteq & 
        \frac{n_{0s}}{\sqrt{\pi}\, v_{\mathrm{th},\parallel s}} 
        \exp \left[-\frac{\left(v_{\parallel} - V_{\parallel s}\right)^2}
        {v_{\mathrm{th},\parallel s}^2}\right] \\
        &\times
        \frac{1}{\pi\, v_{\mathrm{th},\perp s}^2} 
        \exp \left(-\frac{v_{\perp}^2}{v_{\mathrm{th},\perp s}^2}\right).
    \end{aligned}
    \label{eq:maxwell}
\end{equation}

The total distribution function $f(v_{\parallel}, v_{\perp}) = \sum_{s} f_{\mathrm{bi\text{-}M},\, s}$, where the sum runs over all species (electrons, proton core, beam\footnote{Note that \citet{verniero2022} referred to the intermediate population between the core and hammerhead as the ``beam'', whereas more recent work by \citet{bharatidas2026a} refers to it as the ``neck''. In this paper, the original nomenclature of \citet{verniero2022} is adopted.}, and hammerhead), is evolved self-consistently over time by the code.

\begin{table}[ht!]
\centering
\caption{Characteristic quantities recovered from the triple bi-Maxwellian fit of the proton VDF at 2020-01-29/18:16:23 (see \citealt{verniero2022}): thermal and drift velocities, normalized to the proton-core Alfv\'{e}n speed $c_\mathrm{Ap}$, and number densities $n_s/n_0$, expressed as fractions of the total density, for each species: electrons, proton core, proton beam, and proton hammerhead. \label{tab:speed}}
\begin{tabular}{lcccc}
\hline \hline
& Electrons & Core & Beam & Hammerhead \\
\hline
$n_{s}/n_{0}$                                        & 1.00  & 0.80  & 0.15 & 0.05 \\
$v_{\mathrm{th},\parallel s}/c_{\mathrm{Ap}}$        & 4.53  & 0.45  & 1.00 & 7.30 \\
$v_{\mathrm{th},\perp 1, s}/c_{\mathrm{Ap}}$         & 4.53  & 4.70  & 0.99 & 1.47 \\
$v_{\mathrm{th},\perp 2, s}/c_{\mathrm{Ap}}$         & 4.53  & 4.70  & 0.99 & 1.47 \\
$V_{\parallel s}/c_{\mathrm{Ap}}$                    & 0.00  & $-$0.41 & 0.98 & 3.50 \\
\hline
\end{tabular}
\end{table}
\twocolumngrid

The thermal velocities in Equation~\eqref{eq:maxwell} are initialised to the values listed in Table~\ref{tab:speed}. The proton--electron plasma satisfies the quasi-neutrality condition $n_{\mathrm{e}} \approx n_{\mathrm{p}} = n_{\mathrm{c}} + n_{\mathrm{b}} + n_{\mathrm{h}}$ and the zero-net-current condition $n_{\mathrm{c}} V_{\parallel\mathrm{c}} + n_{\mathrm{b}} V_{\parallel\mathrm{b}} + n_{\mathrm{h}} V_{\parallel\mathrm{h}} = 0$, which together ensure that the electromagnetic field is initially free of fluctuations.

Figure~\ref{fig:qlt} compares the unstable eigenmodes as a function of mass ratio; in particular, we compare the realistic ratio $m_i/m_e=1836$ with a reduced one $m_i/m_e=50$. Qualitatively, the spectra show no morphological differences between the two cases, and the most unstable eigenmode (MUE) appears at the same location in the $(k_{\parallel}, k_{\perp})$ wavevector space. The left column (panels~a and~d) shows the real frequency, which exhibits vertical stripes of uniform magnitude across $k_{\perp}$, indicating that $\omega^{\star}$ depends only on $k_{\parallel}$. The fact that $\omega^{\star}/\omega_{\mathrm{pp}} \in \mathbb{R}_{+}$, where $\omega_{\mathrm{pp}}$ is the proton plasma frequency, confirms that the MUE is non-stationary and propagates in the positive parallel direction ($\boldsymbol{k} \times \boldsymbol{B} = \boldsymbol{0}$). The middle column (panels~b and~e) shows the growth rate, where we highlight the peak value $\gamma^{\star}/\omega_{\mathrm{pp}}$.

The right column shows the polarization $\mathcal{P} \doteq \mathbb{R}\!\left\{ i\,\mathrm{sgn}(\omega)\,E_x/E_y \right\}$ \citep{stix1962,gary1993}, where $E_x$ and $E_y$ are the electric field components along $x$ and $y$, respectively. Since $\mathcal{P} \in \mathbb{R}_{+}$ and its magnitude is close to unity, the wave is right-handed (RH) and circularly polarized (CP), rotating counterclockwise in the plane perpendicular to the propagation direction. These properties are the signature of a quasi-parallel mode belonging to the Fast-Magnetosonic/Whistler (FM/W) branch \citep{pezzini2024, pezzini2026}.

Quantitatively, both the real and imaginary frequency spectra (left and middle columns) appear more intense in the reduced mass ratio case, whereas the polarization (right column) remains unchanged. To quantify the discrepancy, we define the relative error $\epsilon(q) \doteq |q_{50} - q_{1836}|/q_{1836}$, where $q$ is an arbitrary quantity. For the real frequency, $\epsilon(\omega^{\star}/\omega_{\mathrm{pp}}) \approx 3\%$, indicating a negligible difference; for the growth rate, $\epsilon(\gamma^{\star}/\omega_{\mathrm{pp}}) \approx 12\%$. In conclusion, reducing the mass ratio primarily enhances the growth rate without significantly distorting the spectral morphology and, therefore, without qualitatively changing the evolution of the physical system. Therefore, a reduced mass ratio $m_{\mathrm{p}}/m_{\mathrm{e}} = 50$ is adopted for numerical convenience. However, when drawing physical conclusions about time-dependent quantities, a rescaling must be applied: temporal quantities are overestimated by approximately $12\%$, while frequencies are underestimated by approximately $3\%$. This is a rough estimate based on the assumption that reducing the mass ratio affects time-dependent quantities linearly.

\section{Numerical setup} \label{sec:setup}
The \textsc{ECsim} code \citep{lapenta2017, bacchini2023, croonen2024} was employed to solve the Vlasov--Maxwell system of equations for the unstable proton--electron plasma described in Section~\ref{sec:linear}, with the distribution defined by Equation~\ref{eq:maxwell} and the parameters listed in Table~\ref{tab:speed}. The simulation was carried out in a two-dimensional periodic Cartesian domain $(x, y)$.

The MUE has a parallel wavenumber $k_{\parallel}/d_{\mathrm{p}}^{-1} \approx 0.83$, where $d_{\mathrm{p}}$ is the proton skin depth; the perpendicular wavenumber is chosen to $k_{\perp}/d_{\mathrm{p}}^{-1}  \approx 0.05$ among the infinite perpendicular values, consistent with the predominantly parallel nature of the unstable eigenmodes. These wavenumbers yield a parallel wavelength $\lambda_{x} = 2\pi/k_{\parallel} \approx 7d_{\mathrm{p}}$ and a perpendicular wavelength $\lambda_{y} = 2\pi/k_{\perp} \approx 125d_{\mathrm{p}}$. The domain lengths are chosen to accommodate multiple parallel oscillations of the dominant mode, resulting in a rectangular box of dimensions $L_{x}/d_{\mathrm{p}} \approx 35$ and $L_{y}/d_{\mathrm{p}} \approx 140$. The numerical grid consists of $N_{x} \times N_{y} = 256 \times 1024$ cells with a uniform spatial increment $\Delta x/d_{\mathrm{p}} = \Delta y/d_{\mathrm{p}} = 0.135$, yielding a resolution of at least four grid points per cyclotron radius. Each species is represented by $64^2$ particles per cell, initially distributed uniformly across the domain.

The temporal scale is estimated from the growth rate of the MUE, $\gamma^{\star}/\omega_{\mathrm{pp}} \approx 3.684 \times 10^{-4}$, which defines a linear timescale $\tau_{\mathrm{lin}} = 2\pi/\gamma^{\star}$. The simulation is run for a total time $t_{\mathrm{tot}} \sim 5\,\tau_{\mathrm{lin}}$, sufficient to capture the full development of the instability through the nonlinear phase and its dynamics. The time step is set to $\Delta t = 0.128\,\omega_{\mathrm{pp}}^{-1} = 0.128 \times 10^{-2}\,\Omega_{\mathrm{cp}}^{-1}$, where $\Omega_{\mathrm{cp}}^{-1}$ is the proton core cyclotron frequency, ensuring accurate resolution of the proton gyro-period for all species and numerical stability of the algorithm.

\section{Results}\label{sec:results}

\subsection{Global Dynamical Phases}\label{subsec:phases}

\begin{figure}[ht!]
\centering
\includegraphics[width=\columnwidth]{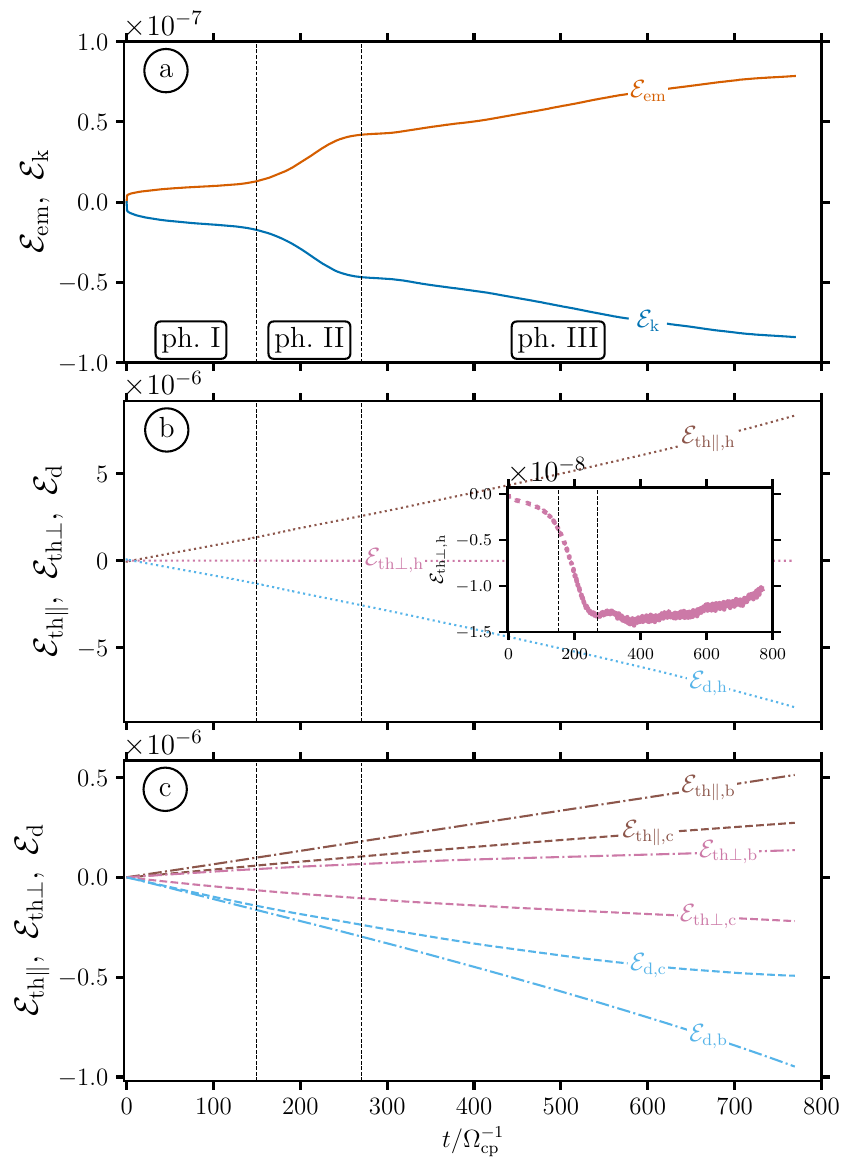}
\caption{Energy diagnostic time series, partitioned into three dynamical phases (ph.~I, ph.~II, ph.~III) separated by vertical dashed lines. (a) Total electromagnetic energy change $\mathcal{E}_{\mathrm{em}}$ (solid blue) and total kinetic energy change $\mathcal{E}_{\mathrm{k}}$ (solid orange). (b) Hammerhead proton energetics (dotted lines): parallel thermal $\mathcal{E}_{\mathrm{th\parallel,h}}$ (brown), perpendicular thermal $\mathcal{E}_{\mathrm{th\perp,h}}$ (purple), and drift $\mathcal{E}_{\mathrm{d,h}}$ (light blue). (c) Core (dashed lines) and beam (dash-dotted lines) proton energetics: analogous color coding as in panel (b).}
\label{fig:energy}
\end{figure}

Figure~\ref{fig:energy} shows the time series of relevant energy quantities integrated over space, providing an overview of the system's global temporal evolution through its different phases. All the plotted energies $\mathcal{E}$ represent the relative change with respect to their initial values $E_{0}$, normalized to the system’s total initial energy $U_{0}$, as $\mathcal{E} = (E - E_{0})/U_{0}$. The system’s electromagnetic energy $\mathcal{E}_{\mathrm{em}}$ and kinetic energy $\mathcal{E}_{\mathrm{k}}$ are defined as
\begin{equation}
    \begin{aligned}
        \mathcal{E}_{\mathrm{em}} &\doteq \int \mathrm{d}^{3}\boldsymbol{x}\, \frac{B^2 + E^2}{8\pi}, \\
        \mathcal{E}_{\mathrm{k}} &\doteq \sum_{s} \left(\mathcal{E}_{\mathrm{th}, s}+ \mathcal{E}_{\mathrm{ d},s}\right).
    \end{aligned}
    \label{eq:kinmag}
\end{equation}
Here, we define the components of $\mathcal{E}_\mathrm{k}$ as follows:
\begin{equation}
    \begin{aligned}
        \mathcal{E}_{{\mathrm{th}}, s} &\doteq \mathcal{E}_{{\mathrm{th}} \perp, s}+ \mathcal{E}_{{\mathrm{th}} \parallel, s}= \int {\mathrm{d}}^{3}\boldsymbol{x} \left(p_{\perp,s} + \frac{p_{\parallel,s}}{2} \right), \\
        \mathcal{E}_{\mathrm{d},s} &\doteq \int {\mathrm{d}}^{3}\boldsymbol{x}\frac{\rho_{s} V_{s}^2 }{2}.
    \end{aligned}
    \label{eq:kinspec}
\end{equation}
In Equation~\ref{eq:presstensor}, the pressure tensor and its perpendicular and parallel components for the generic species $s$ are
\begin{equation}
    \begin{aligned}
        \boldsymbol{P}_{s} &\doteq m_{s} \int \mathrm{d}^3 \boldsymbol{v} (\boldsymbol{v} - \boldsymbol{V}_{s})(\boldsymbol{v} - \boldsymbol{V}_{s}) f_{s}, \\
        p_{\perp, s} &\doteq \boldsymbol{P}_{s} \bdbldot (\mathbb{1} - \hat{\boldsymbol{b}} \hat{\boldsymbol{b}})/2, \\
        p_{\parallel, s} &\doteq \boldsymbol{P}_{s} \bdbldot \hat{\boldsymbol{b}} \hat{\boldsymbol{b}}.
    \end{aligned}
    \label{eq:presstensor}
\end{equation}

Here, $\rho_{s}$ is the proton gyroradius and $\hat{\boldsymbol{b}} = \boldsymbol{B}/\|\boldsymbol{B}\|$ is the unit vector along the magnetic field direction. 

Figure~\ref{fig:energy}(a) displays the time evolution of the two energies from Equation~\eqref{eq:kinspec}, demonstrating the conversion of $\mathcal{E}_{\mathrm{k}}$ into $\mathcal{E}_{\mathrm{e}m}$, driven by the system's evolution. Three distinct phases can be identified, marked by black dashed vertical lines. \textit{Excitation phase} (ph.~I, $0 \lesssim t/\Omega_{\mathrm{cp}}^{-1} \lesssim 150$): the system remains in a metastable equilibrium, perturbed only by the numerical noise inherent in full-kinetic algorithms. The system remains in this state until the perturbations overcome the potential barrier, triggering the instability. \textit{Amplification phase} (ph.~II, $150\lesssim t/\Omega_{\mathrm{cp}}^{-1} \lesssim 280$): the instability develops; $\mathcal{E}_{\mathrm{em}}$ rises rapidly at the expense of $\mathcal{E}_{\mathrm{k}}$, which decreases correspondingly. \textit{Nonlinear phase} (ph.~III, $t/\Omega_{\mathrm{cp}}^{-1} \gtrsim 280$): a secular growth of $\mathcal{E}_{\mathrm{em}}$ continues at the expense of $\mathcal{E}_{\mathrm{k}}$.

Figure~\ref{fig:energy}(b) shows the energetics of the hammerhead protons, the fastest-drifting population, which dominates the system dynamics. $\mathcal{E}_{\mathrm{th\parallel,h}}$ grows monotonically at the expense of $\mathcal{E}_{\mathrm{d,h}}$, which loses energy at a comparable rate, making their evolutions nearly symmetric. In contrast, $\mathcal{E}_{\mathrm{th\perp,h}}$ shows nonnegligible variation during the \textit{amplification phase}, albeit two orders of magnitude smaller than the other energy components. Figure~\ref{fig:energy}(c) shows the energetics of the core and beam protons, which have slower drift speeds and therefore contribute less to the overall energy balance. Here as well, a similar hierarchy is observed in the evolution: $\mathcal{E}_{\mathrm{th\parallel,b}}$ and $\mathcal{E}_{\mathrm{th\parallel,c}}$ 
($\|\mathcal{E}_{\mathrm{th\parallel,b}}\| > \|\mathcal{E}_{\mathrm{th\parallel,c}}\|$) grow at the expense of $\mathcal{E}_{\mathrm{d,b}}$ and $\mathcal{E}_{\mathrm{d,c}}$ ($\|\mathcal{E}_{\mathrm{d,b}}\| > \|\mathcal{E}_{\mathrm{d,c}}\|$), respectively. $\mathcal{E}_{\mathrm{th\perp,b}}$ grows at the expense of $\mathcal{E}_{\mathrm{th\perp,c}}$, but with a smaller parallel-to-perpendicular ratio. Unlike the energetics in Figure~\ref{fig:energy}(a), those in panel~(c) exhibit no clear symmetry, and their rates of change follow no simple pattern. Finally, neither panel~(b) nor panel~(c) displays the growth behaviour observed in ph.~II of panel~(a).

\subsection{Waves and Their Spectral Properties}\label{subsec:wave}

\begin{figure}[ht!]
\includegraphics[width=1\columnwidth]{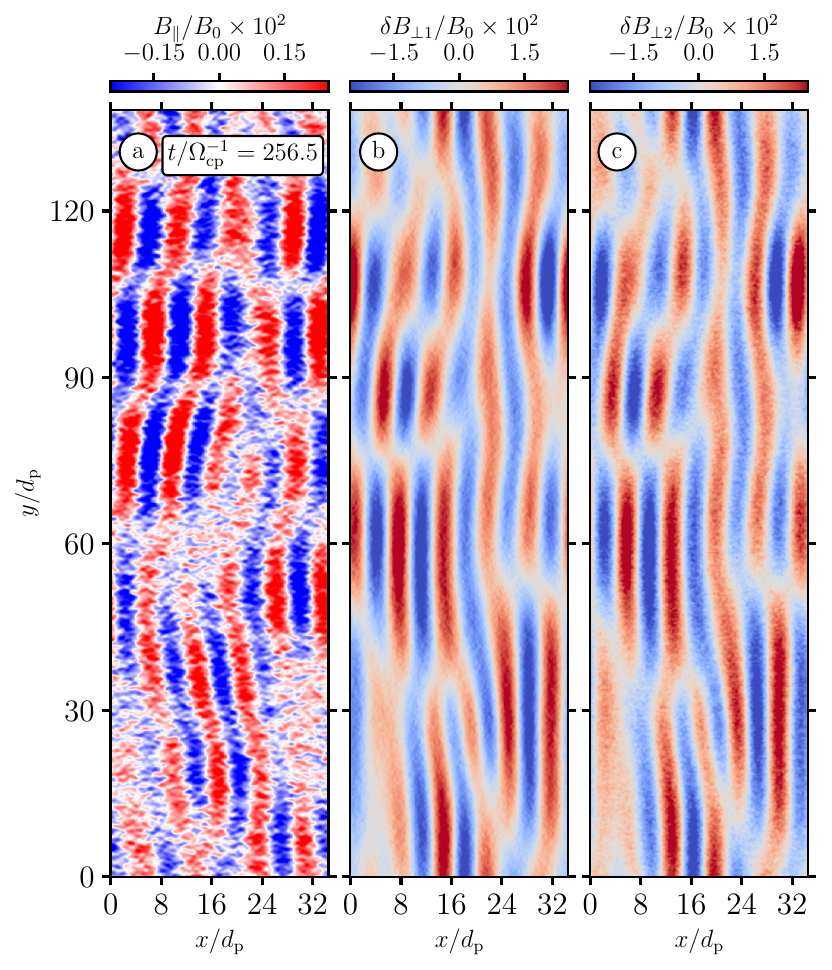}
\caption{Spatial distribution of the electromagnetic field fluctuations at the end of the \textit{amplification phase}. Panel~(a): parallel magnetic fluctuation $\delta B_{\parallel}/B_0$. Panels~(b) and~(c): perpendicular magnetic fluctuation components $\delta B_{\perp 1}/B_0$ and $\delta B_{\perp 2}/B_0$, respectively.}
\label{fig:fields}
\end{figure}

During the \textit{excitation phase}, noise fluctuations are effectively exploited to destabilize the system, seeding the instability which fully develops during the \textit{amplification phase}, where the system strongly amplifies waves. Once wave amplification saturates, the system enters a stage in which waves propagate, nonlinearly interact, and eventually feed back onto the particles, causing heating: this is the \textit{nonlinear phase}.

Figure~\ref{fig:fields} shows the magnetic field fluctuations at the end of the \textit{amplification phase}. Panel~(a) displays $B_\parallel/B_0$, the parallel fluctuation with an amplitude reaching $\sim \pm 0.2\times 10^{-2}\,B_0$, roughly an order of magnitude smaller than the transverse components, suggesting the transverse nature of the wave. The $B_\parallel/B_0$ fluctuations feature wave fronts aligned along $y$; however, the wave pattern becomes disrupted and irregular in localized regions of the domain. In addition, $B_\parallel/B_0$ is correlated with density fluctuations, suggesting the compressible nature of the FM/W modes \citep{pezzini2024}.
Panels (b) and (c) show $\delta B_{\perp 1}/B_0$ and $\delta B_{\perp 2}/B_0$, the transverse fluctuation components that carry most of the wave power, with amplitudes reaching $\sim \pm 2.0\times 10^{-2}\,B_0$. Both panels display coherent, large-scale striations that are predominantly aligned with the $y$-axis but also display some inclination with respect to the $x$-axis, producing wave fronts that propagate slightly obliquely across the domain. 

\begin{figure}[ht!]
\includegraphics[width=1\columnwidth]{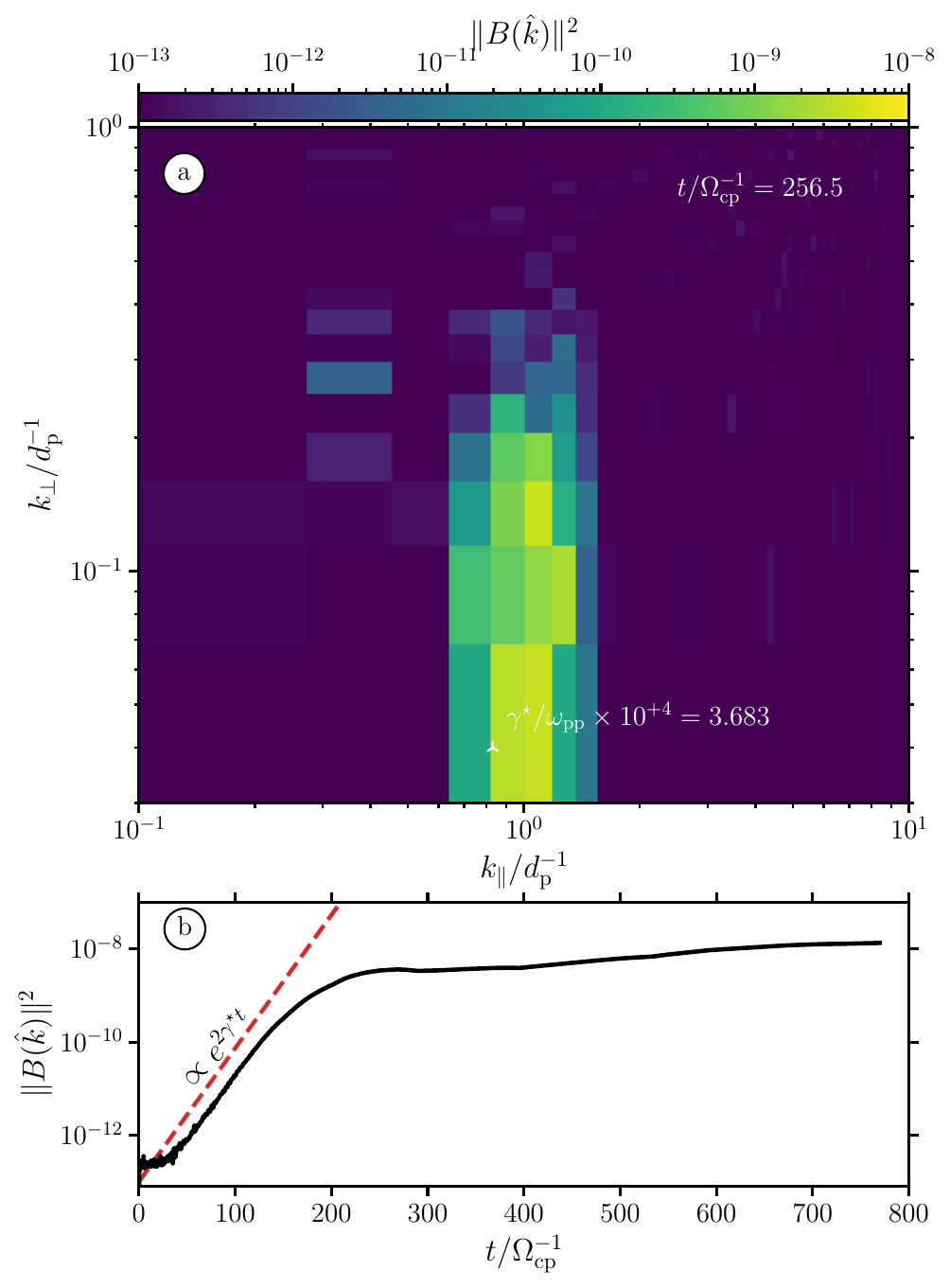}
\caption{Spectral analysis of the unstable eigenmodes. Panel~(a): power spectral density $\|\hat{B}(\hat{k})\|^2$ in the $(k_{\parallel}, k_{\perp})$-space at $t\,\Omega_{\mathrm{cp}}^{-1} = 256.5$. Panel~(b): temporal evolution of the power spectral density at the fastest growing mode. The dashed red line shows the linear theory prediction $\propto e^{2\gamma^{\star} t}$.}
\label{fig:spectra}
\end{figure}

Figure~\ref{fig:spectra} displays the spectral analysis of the fluctuations reported in Figure~\ref{fig:fields}. Panel~(a) shows the power spectral density (PSD) in the $(k_{\parallel}, k_{\perp})$-space, which exhibits a narrow peak concentrated around $k_{\parallel}/d_{\mathrm{p}}^{-1} \approx 1$ across a range of $k_{\perp}$ values, consistent with a quasi-monochromatic parallel unstable eigenmode. The MUE mode is highlighted with the white marker. Panel~(b) shows the temporal evolution of the PSD at this mode, whose exponential growth rate is in agreement with the linear theory prediction $\gamma^{\star}$. The spectral properties are fully compatible with the linear theory results reported in Figure~\ref{fig:qlt}.

\subsection{Proton Velocity Distribution Functions}\label{subsec:vdf}

\begin{figure*}[ht!]
\includegraphics[width=1\textwidth]{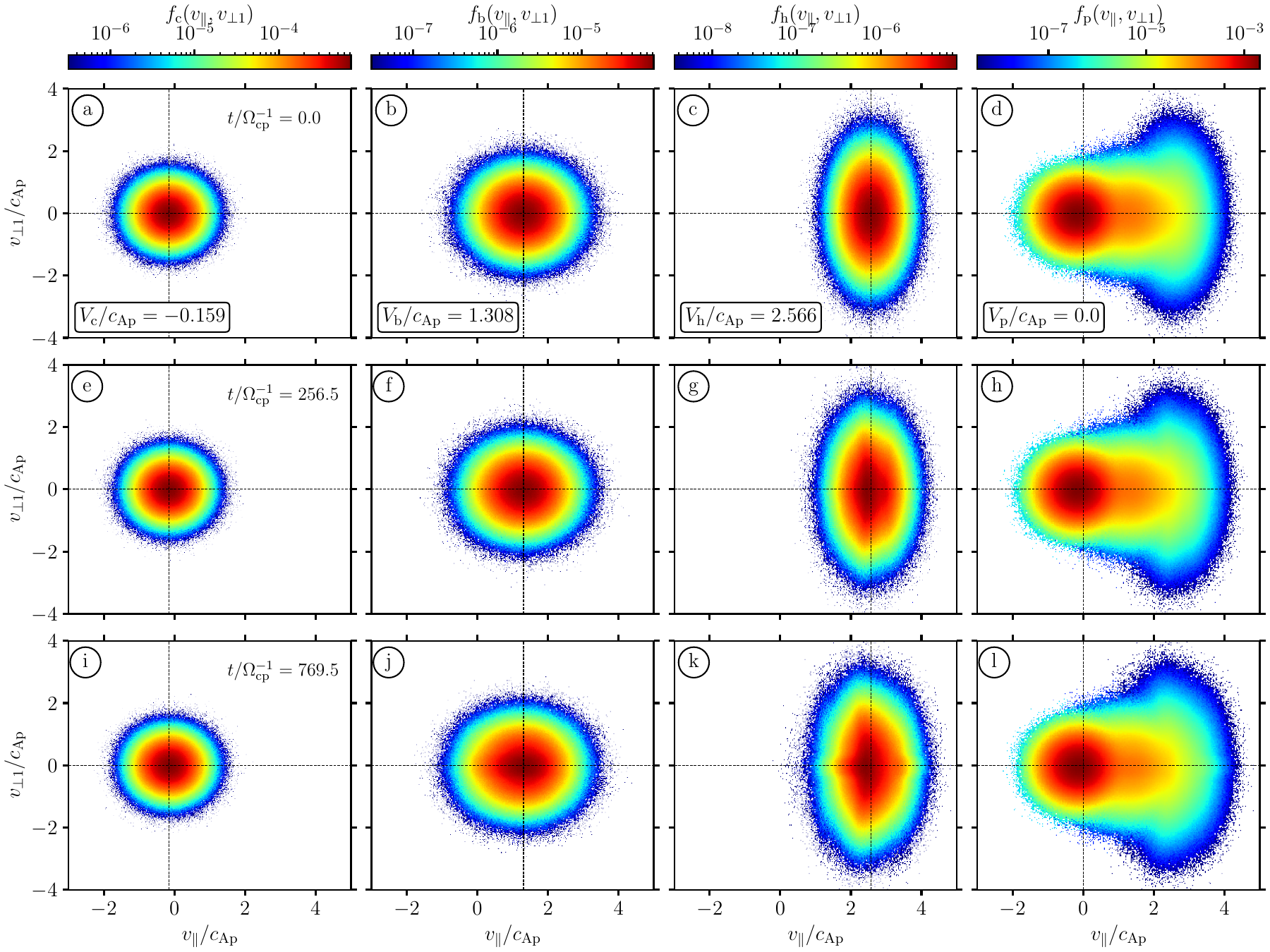}
\caption{pVDFs in $(v_{\parallel}, v_{\perp 1})$-space at three representative times: $t/\Omega_{\mathrm{cp}}^{-1} = 0.0$ (top row, panels (a)--(d)), $256.5$ (middle row, panels (e)--(h)), and $769.5$ (bottom row, panels (i)--(l)), corresponding to the excitation, amplification, and nonlinear phases of the instability, respectively. Columns 1--4 correspond to the core ($f_{\mathrm{c}}$), beam ($f_{\mathrm{b}}$), hammerhead ($f_{\mathrm{h}}$), and total proton ($f_{\mathrm{p}}$) distributions, respectively. Dashed lines mark the initial drift velocity of each population.}
\label{fig:vdf}
\end{figure*}

The proton velocity distribution function (pVDF) is shown for both the individual proton populations and the total proton distribution. Figure~\ref{fig:vdf} displays the evolution of the pVDFs in the reduced $(v_{\parallel}, v_{\perp 1})$-space across the three dynamical phases identified in Figure~\ref{fig:energy}.
During the \textit{excitation phase} ($t/\Omega_{\mathrm{cp}}^{-1} = 0.0$, panels (a)--(d)), all populations retain their prescribed nonequilibrium parameters: anisotropies and initial drift velocities as listed in Table~\ref{tab:speed}. The hammerhead (panel~(c)) already displays a strong positive anisotropy, with a characteristic elongation in $v_{\perp}$ that clearly distinguishes it from the quasi-isotropic core (panel~(a)) and beam (panel~(b)) distributions.
During the \textit{amplification phase} ($t/\Omega_{\mathrm{cp}}^{-1} = 256.5$, panels~(e)--(h)), the core (panel~(e)) and beam (panel~(f)) remain largely unchanged, consistent with their marginal role in driving the instability. The hammerhead (panel~(g)), in contrast, shows a clear rearrangement of its isocontours near $v_{\parallel} = V_{\mathrm{h}}$: the elliptical contours of panel~(c) have been distorted, indicating the onset of wave--particle interactions around $V_{\mathrm{h}}$. The total VDF (panel~(h)) marginally reflects this evolution relative to panel~(d).
During the \textit{nonlinear phase} ($t/\Omega_{\mathrm{cp}}^{-1} = 769.5$, panels (i)--(l)), the core (panel~(i)) remains quasi-isotropic and essentially unchanged, confirming its passive role throughout the instability. The beam (panel~(j)) exhibits only modest parallel broadening near $v_{\perp 1}/c_{\mathrm{Ap}} \approx 0$. The hammerhead (panel~(k)) has evolved further in the parallel direction, with a pronounced bump near $v_{\perp 1}/c_{\mathrm{Ap}} \approx 0$. The total pVDF (panel~(l)) reflects the parallel broadening characteristic of the hammerhead species.

\begin{figure*}[ht!]
\centering
\includegraphics[width=0.90\textwidth]{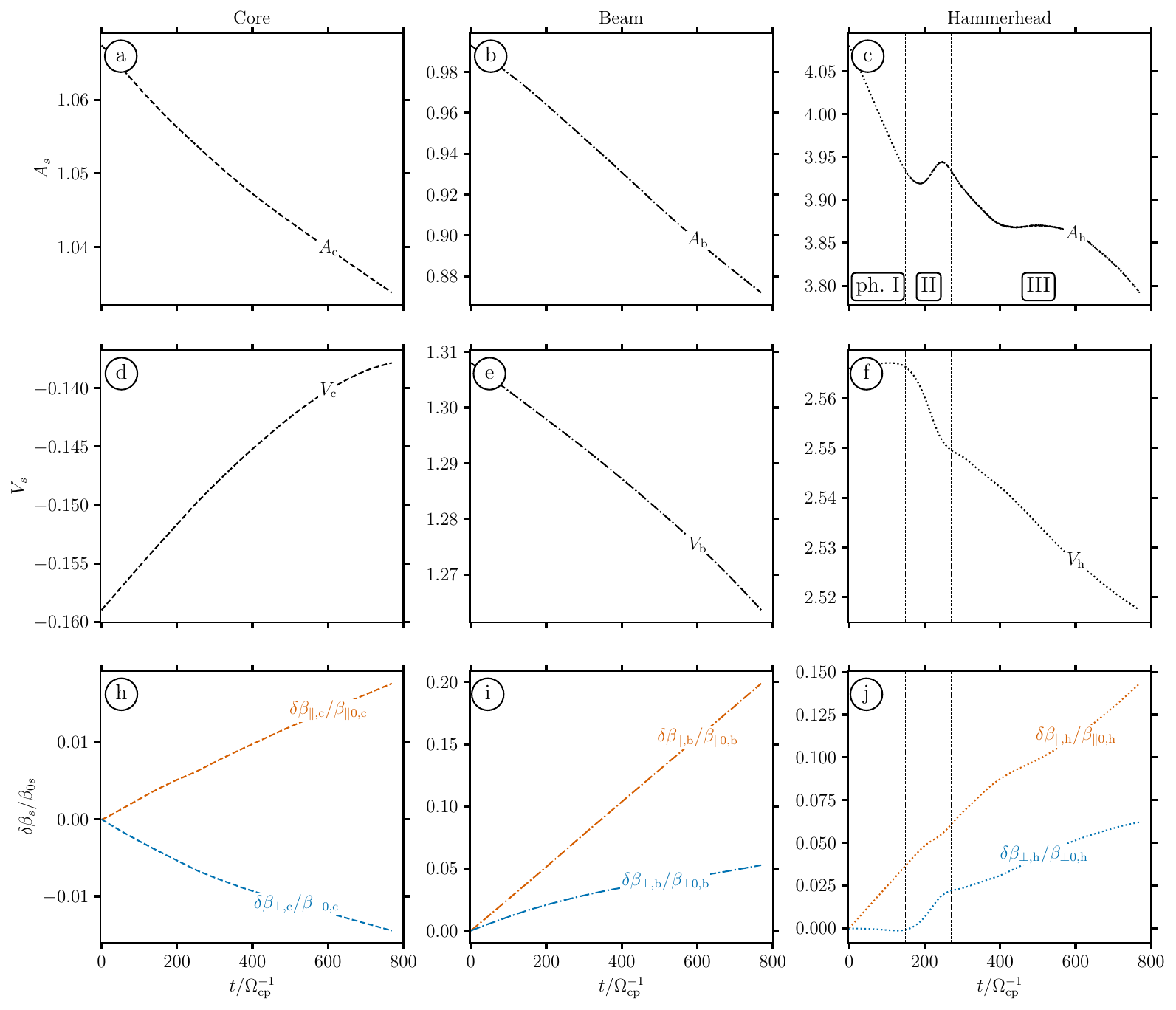}
\caption{Temporal evolution of the kinetic properties of the four proton populations (columns: core, beam, hammerhead, from left to right). Top row, panels~(a)--(c): temperature anisotropy. Middle row, panels~(d)--(f): bulk drift velocity $V_s/c_{\mathrm{Ap}}$ along the background magnetic field. Bottom row, panels~(h)--(j): relative variation of the parallel and perpendicular plasma-beta, $\delta\beta_{\parallel,s}/\beta_{\parallel 0,s}$ (red) and $\delta\beta_{\perp,s}/\beta_{\perp 0, s}$ (blue, where $\delta \beta_s = \beta_s(t) - \beta_{0 s}$ for an arbitrary direction). Vertical dashed lines mark the boundaries between the three dynamical phases (ph.~I,~II,~III) as defined in Figure~\ref{fig:energy}.}
\label{fig:aniso}
\end{figure*}

We analyze the moments of the pVDF as a function of time, focusing on each population individually. Figure~\ref{fig:aniso} shows the temporal evolution of the temperature anisotropy $A_s = T_{\perp,s}/T_{\parallel,s}$, where the temperatures are computed from the pressure tensor as $T_{\perp, s} = p_{\perp, s}/n_{s}$ and $T_{\parallel, s} = p_{\parallel, s}/n_{s}$; bulk drift velocity $V_{s}$; and relative plasma-beta variation $\delta\beta_{s}/\beta_{0,s}$ for the core, beam, and hammerhead populations.
\textit{Core}: (a) $A_{\mathrm{c}}$ is initialized to $\approx 1.068$ and decreases throughout the evolution of the system, reaching $\approx 1.034$ with a variation of $3.15\%$; (d) $V_{\mathrm{c}}/c_{\mathrm{Ap}}$ evolves from $\approx -0.159$ to $\approx -1.138$ with a variation of $13.31\%$; (h) $\delta\beta_{\mathrm{c}}/\beta_{0, \mathrm{c}}$ shows a marginal variation in which $\beta_{\parallel, \mathrm{c}}$ increases and $\beta_{\perp,\mathrm{c}}$ decreases almost symmetrically by around $1\%$; we remark that $\beta_{\perp, s}\propto T_{\perp, s}$ and $\beta_{\parallel, s}\propto T_{\parallel, s}$.
\textit{Beam}: (b) $A_{\mathrm{b}}$ steadily decreases from an initial value of $\approx 0.993$ to a value of $\approx 0.872$ by $12.20\%$; (e) $V_{\mathrm{c}}/c_{\mathrm{Ap}}$ decreases from an initial value of $\approx 1.308$ to a value of $\approx 1.264$ by $3.40\%$; (i) $\delta\beta_{\parallel,\mathrm{b}}/\beta_{\parallel 0,\mathrm{b}}$ exhibits a monotonically increasing and approximately linear profile, while $\delta\beta_{\perp,\mathrm{b}}/\beta_{\perp 0,\mathrm{b}}$ grows more modestly. \textit{Hammerhead}: (c) $A_{\mathrm{h}}$ is initialised to $\approx 4.08$ and rapidly decreases to $\approx 3.91$ during the \textit{excitation phase}, with a variation of $3.80\%$. During the \textit{amplification phase}, it increases abruptly by $0.38\%$ and then decreases to $\approx 3.79$ during the \textit{nonlinear phase} with a total variation of $3.75\%$; (f) $V_{\mathrm{h}}/c_{\mathrm{Ap}} = 2.566$ at initialisation; it then remains almost constant during the \textit{excitation phase}, and drops by $0.62\%$ during the \textit{amplification phase}, then decreases steadily to $2.517$ with a total variation of $1.29 \%$; (j) both $\delta\beta_{\parallel,\mathrm{h}}/\beta_{\parallel 0,\mathrm{h}}$ and $\delta\beta_{\perp,\mathrm{h}}/\beta_{\perp 0,\mathrm{h}}$ grow throughout the simulation, with the parallel component dominating over the perpendicular one. The parallel component grows quasilinearly, while the perpendicular one increases abruptly during the \textit{amplification phase}, then reduces its growth rate during the \textit{nonlinear phase}.

\subsection{Energy Transfer Rate}\label{subsec:fpc}
\begin{figure*}[ht!]
    \centering
    \makebox[\textwidth][c]{\includegraphics[width=1.0\textwidth, trim={2.5cm 0.5cm 2.5cm 0.5cm}, clip]{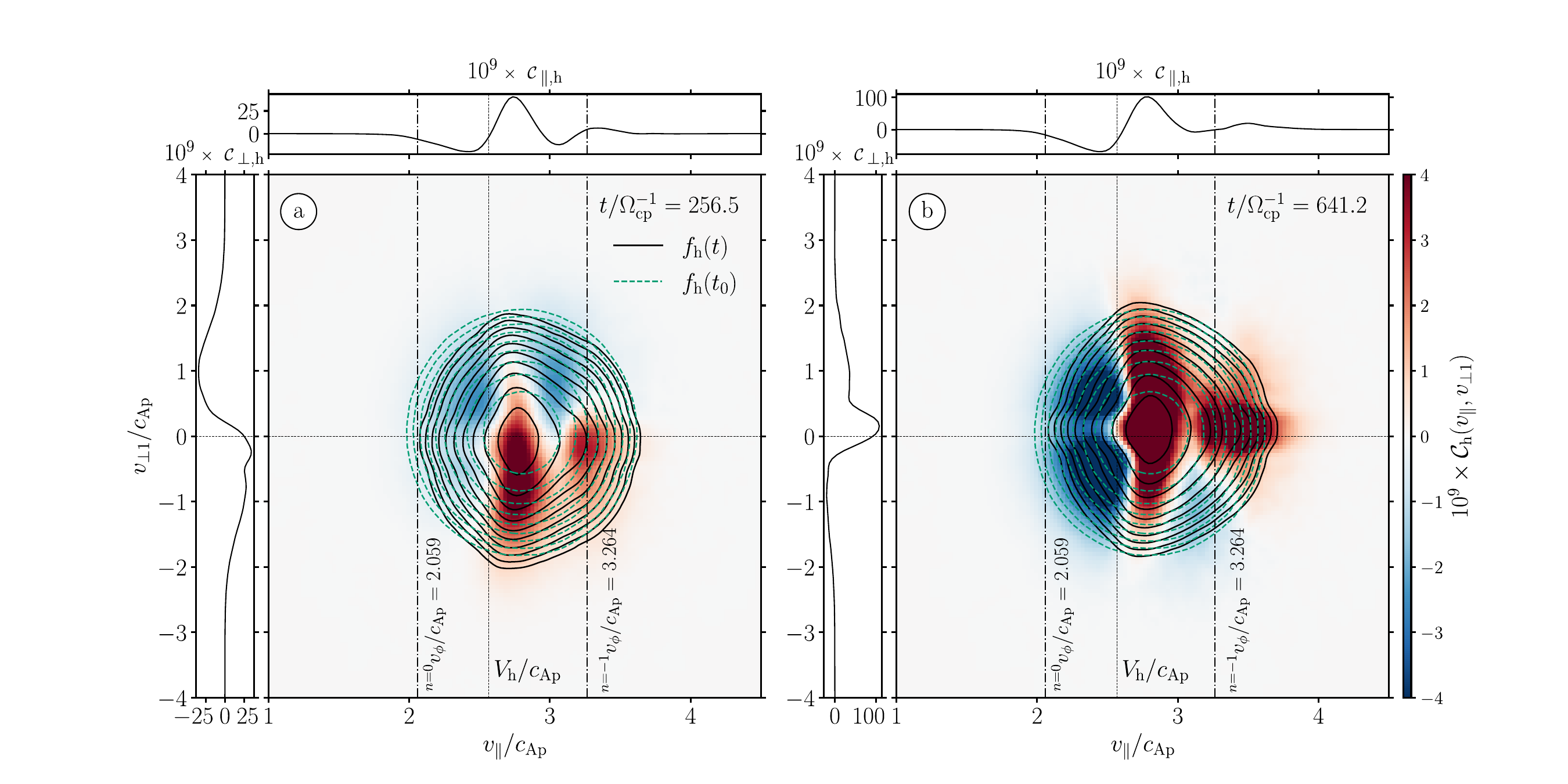}}
    \caption{Field--particle correlation $\mathcal{C}_{\mathrm{h}}(v_{\parallel}, v_{\perp 1})$ for the hammerhead population in the reduced $(v_{\parallel}, v_{\perp 1})$-space at two representative times: $t/\Omega_{\mathrm{cp}}^{-1} = 256.5$ (panel~(a), \textit{amplification phase}) and $t/\Omega_{\mathrm{cp}}^{-1} = 641.2$ (panel~(b), \textit{nonlinear  phase}). In both panels the isocontours show the hammerhead pVDF $f_{\mathrm{h}}(v_{\parallel}, v_{\perp 1})$ at $t/\Omega_{\mathrm{cp}}^{-1} = 641.2$ (solid black line) and at $t/\Omega_{\mathrm{cp}}^{-1} = 0.0$ (dashed green line). The vertical dashed line marks the initial drift velocity $V_{\mathrm{h}}/c_{\mathrm{Ap}} = 2.566$, and the dash-dotted lines mark respectively the $n=0$ Landau resonance and the $n=-1$ cyclotron resonance, with phase speeds ${}^{n}v_{\phi} \doteq (\omega^{\star} - n\Omega_{\mathrm{cp}})/k_{\parallel}^{\star}$. Top panels: parallel marginal correlation $\mathcal{C}_{\parallel \mathrm{h}}(v_{\parallel})$. Left panels: perpendicular marginal correlation $\mathcal{C}_{\perp \mathrm{h}}(v_{\perp 1})$.}
    \label{fig:fpc}
\end{figure*}

The field--particle correlation (FPC) technique \citep{klein2016, howes2017, klein2017} is employed to identify which mechanism extracts energy from the electromagnetic fluctuations (e.g., resonant, nonresonant, or intermittent dissipation) and irreversibly transfers it to the plasma as heat. More precisely, it quantifies the energy transfer rate (ETR) between waves and particles; since multiple processes compete simultaneously, the FPC naturally provides a hierarchy of dissipation channels, making it easier to identify the dominant one.

The aim is to quantify the power density $\mathcal{W}_{s}$ (units: $\mathrm{erg\,cm^{-3}\,s^{-1}}$, i.e., energy per unit volume per unit time) associated with the particle species $s$, where $\mathcal{W}_{s} \propto \boldsymbol{J} \cdot \boldsymbol{E}$. However, both $\boldsymbol{J}$ and $\boldsymbol{E}$ are heavily affected by numerical noise in PIC simulations. Following the approach described in \cite{jiang2024}, the phase-space energy density is instead expressed directly in terms of the pVDF,
\begin{equation}     
    \mathcal{W}_{s}\left(v_{\parallel}, v_{\perp 1}, t\right) =      
    \frac{1}{2} m_{s} v^2 f_{s}\left(v_{\parallel}, 
    v_{\perp 1}, t\right),     
    \label{eq:energydensity} 
\end{equation}
where $v = (v_{\parallel}^2 + v_{\perp 1}^2)^{1/2}$ is the particle speed. The corresponding two-dimensional ETR is then
\begin{equation}     
    \mathcal{C}_{s}\left(v_{\parallel}, v_{\perp 1}\right) = 
    \frac{\partial \mathcal{W}_{s}\left(v_{\parallel}, 
    v_{\perp 1}, t\right)}{\partial t}.     
    \label{eq:fpc} 
\end{equation}

The energy transfer between fields and particles consists of two distinct components: an oscillatory component, associated with undamped wave motion, and a secular component, associated with damped fluctuations. Since only the latter is of physical interest, time averaging over an appropriately chosen correlation interval is typically employed to suppress the oscillatory contribution. Equation~\eqref{eq:fpc} is therefore evaluated using a finite-difference time-domain approximation with a sampling interval $\Delta \tau_{\mathrm{samp}} \approx \tau_{\mathrm{lin}}$ chosen to resolve the secular variations that permanently mark the pVDF. 
Partial integration over the perpendicular and parallel directions yields the corresponding one-dimensional ETRs: 
 \begin{equation}
    \mathcal{C}_{\parallel s}\left(v_{\parallel}\right) = 
    \int_0^{+\infty} 2\pi v_{\perp}\, \mathcal{C}_{s}
    \left(v_{\parallel}, v_{\perp 1}\right) \mathrm{d} v_{\perp 1},
    \label{eq:transferpar}
\end{equation}
\begin{equation}
    \mathcal{C}_{\perp 1 s}\left(v_{\perp}\right) = 
    \int_0^{+\infty} 2\pi v_{\parallel}\, \mathcal{C}_{s}
    \left(v_{\parallel}, v_{\perp 1}\right) \mathrm{d} v_{\parallel},
    \label{eq:transferperp}
\end{equation}
which are equivalently interpretable as the parallel and perpendicular moments of $\mathcal{C}_{s}$, respectively. Finally, to suppress spurious small-scale fluctuations in velocity space, $\mathcal{C}_{s}(v_{\parallel}, v_{\perp 1})$ is convolved with a Gaussian filter during postprocessing.

In general, three cases can be distinguished: $\mathcal{C}_{s} > 0$ defines a phase-space region of net particle accumulation, termed the \textit{pile-up region}; $\mathcal{C}_{s} < 0$ defines a region from which particles are diffused away, termed the \textit{depletion region}; and $\mathcal{C}_{s} = 0$ defines the \textit{neutral region}, in which no net particle diffusion occurs. 


To track particle fluxes in phase space, pVDF values are compared between the initial state and the time of interest, with the difference quantified by $\Delta_{s} = f_{s}(t)-f_{0 s}$. Analogously to $\mathcal{C}_{s}$, three cases arise: where the distribution is positively distorted ($\Delta_{s} > 0$), we have a \textit{pile-up region}; where the distribution is negatively distorted ($\Delta_{s}< 0$), a \textit{depletion region}; and where the distribution is unchanged ($\Delta_{s}= 0$), a \textit{neutral region}.

Figure~\ref{fig:fpc} displays the FPC analysis applied to the hammerhead population. Panel~(a) shows the late amplification phase ($t/\Omega_{\mathrm{cp}}^{-1} = 256.5$): the FPC pattern is asymmetric, displaying a weak double-banded signature consisting of a \textit{depletion region} and a \textit{pile-up region}, with the interface located near $v_{\parallel} \approx V_{\mathrm{h}}$, which is at a higher speed than the expected Landau resonance ${}^{0}v_{\phi}$ calculated from linear theory. Moreover, in the range ${}^{0}v_{\phi} < v_{\parallel} < V_{\mathrm{h}}$ near $v_{\perp}/c_{\mathrm{Ap}} \approx 0$, $\Delta_{\mathrm{h}} \approx 0$ (in white), despite an apparent accumulation of particles being visible. At higher parallel velocities, in correspondence with the anomalous cyclotron resonance at $v_{\parallel} \approx{}^{-1}v_{\phi}$, a secondary \textit{pile-up region} is visible, confined near $v_{\perp 1} \approx 0$ and surrounded by a \textit{depletion region} at higher perpendicular velocities. The parallel moment $\mathcal{C}_{\parallel \mathrm{h}}$ (top panel) shows a sign inversion (from negative to positive) centred around $v_{\parallel} \approx V_{\mathrm{h}}$, and a secondary bump around $v_{\parallel} \approx {}^{-1}v_{\phi}$, while $\mathcal{C}_{\perp \mathrm{h}}$ remains asymmetric. The perpendicular moment (left panel) shows a sign inversion centred around $v_{\perp 1} \approx 0$.

Panel~(b) shows the nonlinear phase ($t/\Omega_{\mathrm{cp}}^{-1} = 641.2$): the pattern observed in panel~(a) is significantly more pronounced and markedly more symmetric. The bipolar structure is now well-consolidated, with a prominent \textit{depletion region} at $v_{\parallel} \lesssim V_{\mathrm{h}}$ extending across all perpendicular velocities. This renders the phase-space region near $v_{\perp 1} \approx 0$ seemingly more prominent.
The corresponding \textit{pile-up region} (red) at $v_{\parallel} \gtrsim V_{\mathrm{h}}/c_{\mathrm{Ap}}$ is intensified accordingly, with the interface between the two regions coinciding with the drift velocity $v_{\parallel} \approx V_{\mathrm{h}}$ (dashed vertical line). The secondary \textit{pile-up region} near $v_{\parallel} \approx {}^{-1}v_{\phi}$, confined close to $v_{\perp 1} \approx 0$, is also more intense compared to panel~(a) and now displays a surrounding halo toward higher $v_{\perp 1}$, indicative of diffusion spreading. This phase-space diffusion is also visible through the pronounced offset between the black and green contours, indicating a significant reshaping of $f_{\mathrm{h}}$ with respect to $f_{0 \mathrm{h}}$. Both features are clearly resolved in $\mathcal{C}_{\parallel \mathrm{h}}$ (top panel), which now exhibits a larger amplitude bipolar signature, roughly four times greater than panel~(a) for the main peak; the secondary higher parallel speed peak here appears to be an increase with respect to the zero level of $\mathcal{C}_{\parallel \mathrm{h}}$; $\mathcal{C}_{\perp \mathrm{h}}$ is now quasi-symmetrically peaked around zero, indicating that perpendicular diffusion occurs mostly towards $v_{\perp 1} \approx 0$.

\section{Discussion \& Conclusion}\label{sec:discludion}

We present the first comprehensive stability analysis of the hammerhead velocity distribution through linear theory and nonlinear PIC simulations. The hammerhead has been analysed using the \texttt{DIS-K} linear solver, which demonstrates that the system is unstable to the FM/W eigenmode propagating parallel to the background magnetic field ($\boldsymbol{k} \times \boldsymbol{B} = \boldsymbol{0}$). This wave is an RHCP wave that can interact with protons through anomalous cyclotron interaction \citep{pezzini2024}. Furthermore, linear theory justifies the use of a reduced mass ratio $m_{\mathrm{p}}/m_{\mathrm{e}} = 50$, since this does not dramatically alter the most-unstable eigenmode \citep{narita2017, verscharen2020}.
The nonlinear regime of the hammerhead distribution is then explored using \textsc{ECsim} PIC simulations.

\textit{General Dynamics \& Heating}---The reservoir of free energy of the system resides mostly in the nonzero drift speed of the different species, which makes this system drift-unstable. The different species drift at different bulk speeds, maintaining current neutrality. At initialisation, the plasma is homogeneous and in unstable equilibrium, with a uniform guide field and no external forces; free energy is stored in the relative drift between ion components and in their thermal anisotropy. The system abruptly evolves by reducing the drift speeds of the constituent species, with a net conversion of drift kinetic energy into magnetic energy, driving the amplification of FM/W waves. These waves eventually feed back into the plasma through wave--particle interactions, heating the plasma and driving it toward a metastable equilibrium. The parallel thermal energy increases for all species, most notably for the beam particles, whose parallel temperature increases by $20\%$; for the hammerhead, the increase is by a factor $15\%$.

\textit{Wave--particle Resonant Interactions}---From the onset of the amplification phase, the hammerhead population exhibits a subtle modification in the morphology of the kinetic shells of the pVDF, which develops during the late amplification phase and consolidates itself in the nonlinear phase. The change is clearly visible: The elliptical symmetry of the kinetic shells progressively acquires a distorted shape; the field--particle correlation technique is employed to obtain a more detailed characterization of the process.

Examining the pVDFs from the late amplification phase onward, the hammerhead population appears to undergo a resonant wave--particle interaction, with the FPC displaying a strong double-banded signature characteristic of Landau damping at $v_{\parallel} \lesssim V_{\mathrm{h}}$. In this scenario, particles in the \textit{depletion region} experience a parallel electric field that accelerates them, driving diffusion toward higher parallel velocities. However, the resonance speed predicted by linear theory, ${}^{0}v_{\phi}$, does not precisely coincide with the interface of the double-banded structure, which instead occurs at a parallel velocity $\sim 17\%$ higher than predicted.
At higher parallel velocities, around $v_{\parallel} \approx {}^{-1}v_{\phi}$, a secondary resonant structure is visible. Particles appear to be drawn from a \textit{depletion region} at higher perpendicular velocities and diffused toward lower perpendicular velocities, accumulating in a \textit{pile-up region} near $v_{\perp 1} \approx 0$, consistent with a cyclotron resonant interaction, which, in this case, is well-aligned with the predicted resonance velocity ${}^{-1}v_{\phi}$.
The fact that $\mathcal{C}_{\mathrm{h}}$ is growing in magnitude, which is particularly visible for the parallel moments, means that the system is more efficient at damping the waves. We therefore identify these processes as fundamentally nonlinear resonant interactions, consistent with the fact that this phenomenon manifests predominantly at nonlinear timescales.

The misalignment between the predicted and observed Landau resonance velocity may be attributed to several factors, the most significant of which are related to the modeling of the pVDFs and the assumptions underlying linear theory. First, the pVDF in our PIC simulation, as well as \texttt{DIS-K} linear solver, is modeled as a superposition of bi-Maxwellians centered at specific drift velocities instead of a continuous complex VDF as could be assumed from observational data; the misalignment of the Landau resonance may therefore be partly due to the overlapping contributions of the different species (beam and hammerhead in this case). Second, linear theory assumes the VDFs to be static in time, whereas in the simulation, they evolve self-consistently with the magnetic fluctuations, which are additionally subject to numerical noise and errors. A natural zeroth-order step to mitigate the first two issues would be to perform the linear analysis using an arbitrary distribution function solver such as \textsc{ALPS} \citep{verscharen2018}, which accepts arbitrary gyrotropic equilibrium distributions, and to compare the results with those of \texttt{DIS-K}.

\textit{Hammerhead Origin Site}---The statistical study by \cite{bharatidas2026a}, which considered all available PSP encounter data in which the hammerhead distribution was observed, reveals a striking connection with crossings of the HCS. The study analyzed all PSP encounter data collected, including during solar maxima, when the HCS is sufficiently warped to produce near-orthogonal crossings. The probability distribution of observing the hammerhead peaks at $\Phi = 1^{\circ}$ in longitude \citep{bharatidas2026, bharatidas2026a}, distributed symmetrically about the boundary layer of the crossing. The corresponding perpendicular distance to the HCS is $\ell = L\sin\Phi \approx L\Phi$ for small angles, where $L = 10\,R_{\odot}$ is the heliocentric distance. Under the hypothesis that the hammerhead is produced at the HCS, we estimate the timescale over which it is advected away by the SW flow, assumed to propagate at $v_{\mathrm{sw}} \sim 100\,\mathrm{km\,s^{-1}}$, yielding $\tau_{\mathrm{cs}} = \ell / v_{\mathrm{sw}} \simeq 100\,\mathrm{s}$. This is consistent with the hammerhead being a long-lived structure, as it does not violently distort its phase-space morphology. We estimate its lifetime to be on the order of the linear growth timescale, $\tau^{\ast}_{\mathrm{lin}} = 2\pi/\gamma^{\ast} = 528\,\Omega_{\mathrm{cp}}^{-1}$, where $\gamma^{\ast}$ is the growth rate calculated with the realistic mass ratio. This, for typical inner-heliospheric cyclotron frequencies $\Omega_{\mathrm{cp}} \approx 1\,\mathrm{Hz}$, yields $\tau^{\ast}_{\mathrm{lin}} \approx 500\,\mathrm{s}$. Since $\tau_{\mathrm{cs}} \lesssim \tau^{\ast}_{\mathrm{lin}}$, the hammerhead can survive long enough to be advected to the observed angular distance before significantly evolving. The fact that the observed distribution peaks sharply at $\Phi = 1^{\circ}$ may be related to the presence of plasma mixing flows near the HCS boundary, as discussed in \cite{krasnoselskikh2023}. This distance may correspond to the region where plasma from the upstream region and plasma from the proton diffusion region at the reconnection site mix and are subsequently sampled by PSP, resulting in the hammerhead distribution being the dominant feature. The hammerhead could be a proton crescent arising from energisation at a magnetic reconnection site: the field reversal causes the particles to demagnetise and follow meandering (Speiser) orbits, which accelerate them, so that they ultimately form a super-Alfv\'enic beam \cite{hesse2014, bessho2016, burch2016, palmroth2023}. This possibility will be tested in future numerical work.

In conclusion, hammerhead proton VDFs are frequently observed in the inner heliosphere, and as demonstrated by our stability analysis, are marginally stable structures. Rather than violently transitioning to entirely different states, they evolve through nonlinear relaxation via wave--particle resonant interactions, with a dominant Landau damping component and a secondary cyclotron damping contribution, which progressively reshapes the pVDF into a distorted configuration of the kinetic shells. However, their marginal stability may explain the ubiquity of hammerhead distributions in inner-heliospheric observations, opening the possibility that they are produced at the heliospheric current sheet through reconnection and subsequently advected outward to the point of measurement by PSP, consistent with their estimated survival timescale. In addition, the heating of the hammerhead population could be tested observationally in future work using \textsc{hampy}.

This study confirms the necessity of employing fully kinetic particle-in-cell simulations combined with advanced analysis tools to capture the nonlinear signatures of plasma heating. Furthermore, this study motivates a rethinking of the methodology used to study plasma instabilities: rather than initialising the simulation directly with an unstable VDF, a more effective and physically consistent approach consists of driving the waves via a Langevin antenna \citep{tenbarge2012, tenbarge2014, arro2025, arro2026} and using the measured VDF as a benchmark, in line with the framework in which magnetic energy cascades from large to small scales. This approach is further justified by the higher precision of the FIELDS wave measurements relative to SWEAP, as reflected in the instrumental uncertainties of the two suites.


\section*{Acknowledgements}\label{sec:acknow}
L.P. acknowledges support from the Research Foundation Flanders (FWO) PhD fellowship (grant no.~\href{https://app.dimensions.ai/details/grant/grant.13861380}{11PCB24N}). Computational resources and services were provided by the Vlaams Supercomputer Centrum (VSC) (grant no.~2025-15), funded by FWO and the Flemish Government--Department of Economy, Science and Innovation (EWI). L.P. gratefully acknowledges Samuel Badman, Samuel Fordin, Maria Elena Innocenti, Kristopher Klein, Leon Ofman for fruitful discussions.
F.B. acknowledges support from the FED-tWIN programme (profile Prf-2020-004, project ``ENERGY'') funded by the Belgian Federal Science Policy Office (BELSPO), as well as from the FWO Junior Research Project (grant no.~G020224N).
A.N.Z. thanks the Belgian Federal Science Policy Office (BELSPO) for the provision of financial support within the framework of the PRODEX Programme of the European Space Agency (ESA) under contract number 4000147286.
R.A.L. acknowledges support from the Agencia Nacional de Investigación y Desarrollo (ANID), Chile, through FONDECyT (grant no.~1251712).


\section*{Author Contributions}\label{sec:contrib}
L.P.: Conceptualisation; Investigation; Methodology; 
Visualisation; Writing – original draft; Writing – review \& editing. 
J.V.: Observations; Methodology; Writing – review \& editing. 
F.B.: Supervision; Software (\textsc{ECsim--RelSIM}); Writing - review \& editing. 
A.N.Z.: Supervision; Writing - review \& editing. 
G.A.: Writing - review \& editing. 
R.A.L: Software (\texttt{DIS-K}); Writing - review \& editing.


\section*{Software}\label{sec:software} 
Simulations were performed with the energy-conserving semi-implicit \texttt{ECsim} PIC code \citep{lapenta2017, bacchini2023}. Linear theory results were obtained using the \texttt{DIS-K} linear solver \citep{lopez2021,lopez2023}. All data analysis presented in this work was carried out using the \textsc{Python} programming language \citep{vanrossum1995}, with core libraries including \textsc{NumPy} \citep{harris2020} for numerical computations, \textsc{Matplotlib} \citep{hunter2007} for data visualisation, and \textsc{SciPy}\citep{virtanen2020} for filtering and curve fitting.


\section*{Data Availability}\label{sec:data}
For the sake of transparency and reproducibility, the dataset supporting the findings of this work has been deposited in a Zenodo repository \dataset[doi:10.5281/zenodo.20296807]{https://doi.org/10.5281/zenodo.20296807}. A more extensive version of the dataset will be made available upon reasonable request to the corresponding authors.

\newpage
\bibliography{References}{}
\bibliographystyle{aasjournal}
\end{document}